\documentclass[11pt]{article}
\usepackage{moriond,epsfig}

\def\be{\begin{equation}}
\def\ee{\end{equation}}
\def\bea{\begin{eqnarray}}
\def\eea{\end{eqnarray}}

\begin{document}
\vspace*{4cm}
\title{Heavy Sterile Neutrinos and Neutrinoless Double Beta Decay  }

\author{ Manimala Mitra$^a$, Goran Senjanovi\'c $^b$, Francesco Vissani $^a$  }

\address{$^a$ INFN, Laboratori Nazionali del Gran Sasso, Assergi, Italy, 
$^b$ ICTP, Trieste, Italy}

\maketitle\abstracts{Sterile neutrinos of  mass up to a  few tens of TeV can saturate the present experimental
bound of neutrinoless double beta decay process.  Due to the updated nuclear matrix elements, the bound on mass and 
mixing angle is now  improved by one order of magnitude. We have performed a detailed analysis of neutrinoless double 
beta decay for the  minimal  Type I seesaw scenario. We have shown that in spite of the naive expectation that   
the light neutrinos  give  
the dominant contribution, sterile neutrinos can  saturate the 
present experimental bound of neutrinoless double beta decay process. However, 
in order to be consistent with radiative stability of light neutrino masses, 
the mass scale of sterile neutrinos should be less than $10 \,\rm{GeV}$.
} 

\section{Introduction:}
Neutrinoless double beta decay ($0 \nu 2\beta$)  is a very  important probe of lepton number violation. The process is 
$(A, Z) \to (A, Z+2)+ 2e^-$, where lepton number is violated by two units. On the experimental
side, there is a  very lively situation as far as present and future experiments \cite{epj,expp} are concerned \footnote{
See the exhaustive list of references in \cite{mfg,werner-rev} for the present and future experiments on $0\nu 2\beta$. See 
\cite{werner-rev,bookk,zuber} and list of references in \cite{mfg} for  reviews on $0 \nu 2\beta$. See
 \cite{visreview,rev-ponte,review-osc-petcov,mik-shap,gelmini-rev,kail,nir-garcia-rev,rev-moha-smir,Senjanovic:2011zz} for the reviews on neutrino
physics.}, and 
even a claim of observation of neutrinoless double beta decay \cite{Klapdor} by Klapdor and collaborators. 
The observation of lepton-number 
violating processes would be a cogent manifestation of incompleteness of the standard model, and could be even 
considered as a step toward the understanding of the origin of  the matter. Indeed, this process can be 
described as creation of a pair of electrons in a nuclear transition.

The  exchange of virtual, light neutrinos is a plausible mechanism \cite{0nu2beta-old}  of 
neutrinoless double beta decay process, 
provided they have Majorana mass \cite{Majorana}. However,  if the new physics scale is not too high, 
alternative possibilities may exist,  where  neutrinoless double beta decay 
is mostly due to mechanisms different from the light neutrino exchange. 
Infact this possibility has been proposed since long \cite{feinberg} and has been widely discussed in 
the  literatures \cite{ms0nu2beta,pion-ex,tello,ibarra-blenow,choubey-sruba}
(see the other references in \cite{mfg}). With this motivation in mind, we have 
considered the minimal Type I seesaw \cite{seesawall}  scenario, and 
we have analyzed how sterile neutrinos can give dominant contribution in neutrinoless double beta decay process. 
We discuss the following points in this context:\\
\\
  i) Large contribution from light neutrino states and constraints from cosmology.\\
 ii) Contribution from sterile neutrino states and   bound on the active-sterile mixing.\\
iii) Naive expectations from the  sterile neutrino states in Type I seesaw.\\
 iv) Possibility of dominant sterile neutrino contribution in Type I seesaw.\\
 %,  that is not suppressed by the smallness of neutrino mass scale.\\
  v) Upper bound on  the mass scale of sterile neutrinos.\\

\section{Light neutrino contribution and constraints from cosmology}

If the light neutrinos are Majorana particle, they can mediate the neutrinoless double beta decay 
process.  The observable is the $ee$ element of the light neutrino mass matrix $|m_{ee}|$ (also denoted by  'effective mass'), where 
$|m_{ee}|=|   \; \sum_i U_{ei}^2\; m_i\; |$, $U$ is the PMNS mixing matrix. Certainly 
$m_{ee}$ is smaller than  $\sum_{i} |U_{ei}^2| m_i$. 
From cosmology, we have bound on the  sum of light neutrino masses {\it i.e.,}
$m_{\mbox{\tiny cosm}} =\sum_{i} m_i$ . For the lightest neutrino mass scale   $m_{\mbox{\tiny min}}>0.1$ eV, relevant 
to the case of present 
experimental sensitivities, one can approximate $|m_{ee}|<  m_{\mbox{\tiny cosm}}/3\approx m_{\mbox{\tiny min}}$.

\begin{figure}[htb]
\begin{center}
\epsfig{file=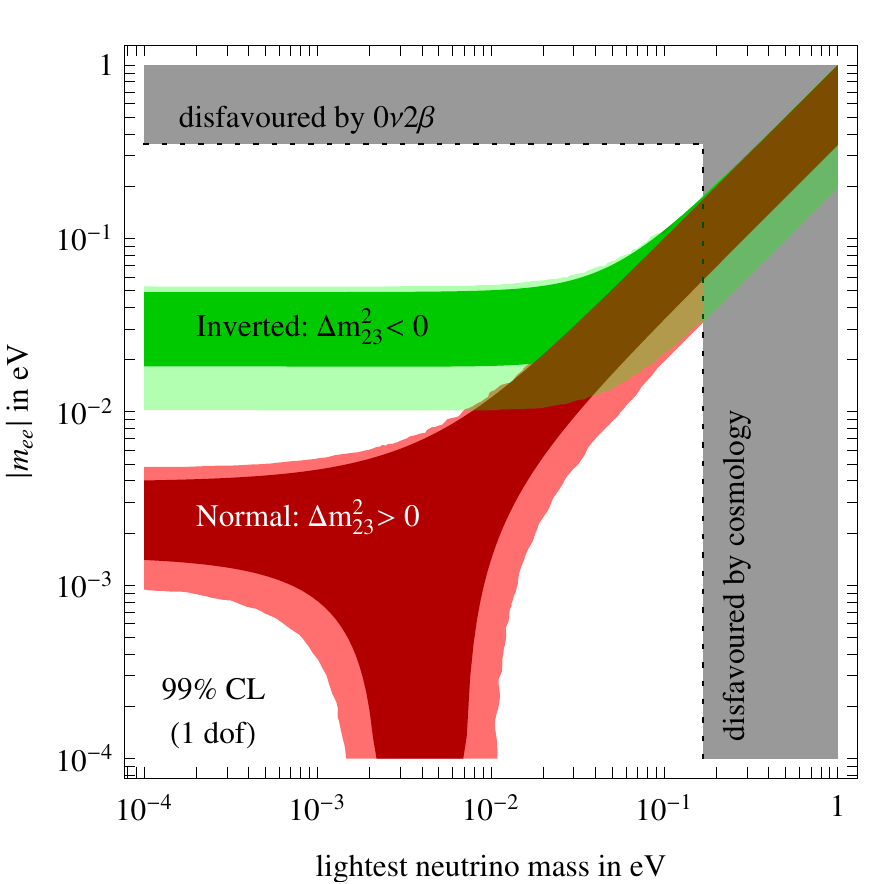,width=0.30\textwidth}
\hspace*{2.5cm}
\epsfig{file=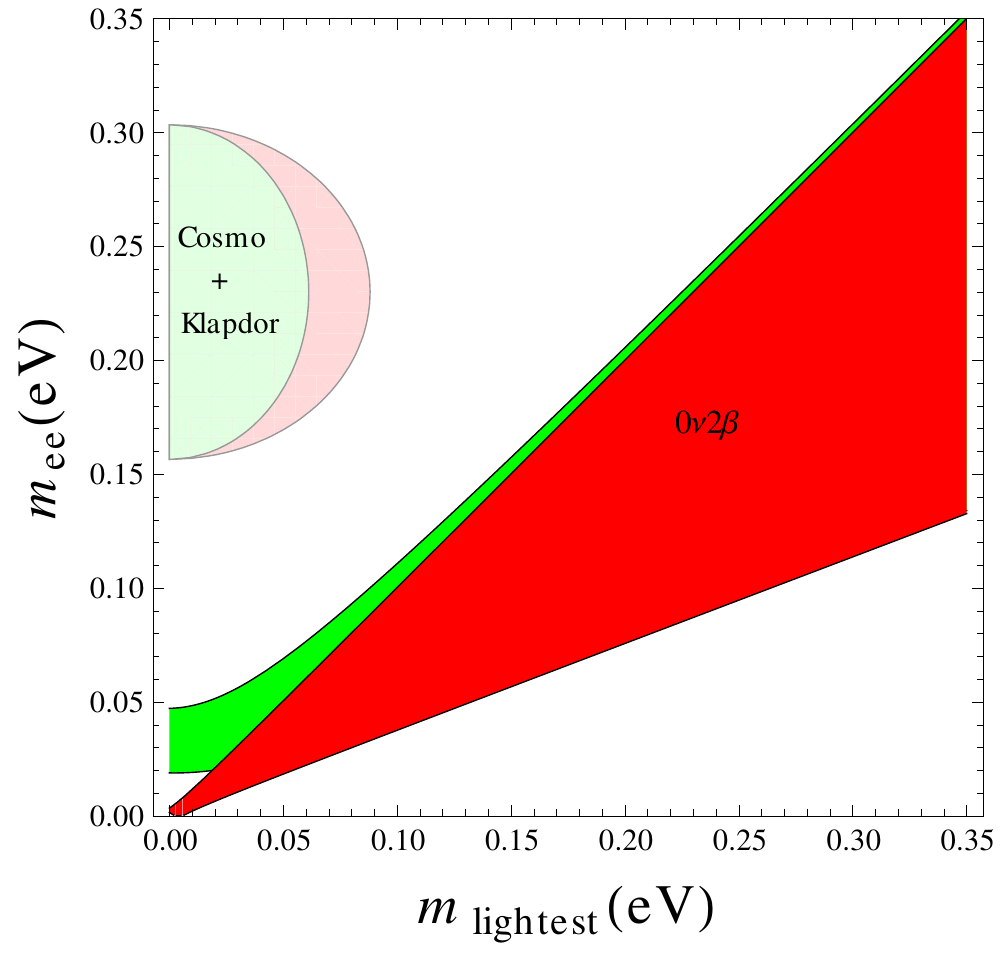,width=0.30\textwidth}
\end{center}
\caption{In the left panel, the  red and green regions represent the effective light neutrino mass for normal and inverted hierarchy. In the right panel, the same but using a linear scale to emphasize the
region presently under
study. Moreover, we show  in light colors the regions resulting from the
combination of the  recent cosmological bound  and Klapdor’s claim at 
$95 \%$ \it{C.L.}}
\label{cosmo-klap}
\end{figure}

The bound coming from 
Heidelberg-Moscow experiment \cite{epj} is  $T_{1/2}>1.9\times 10^{25} \, \rm{yrs}$ at $90\%$ {\em{C.L}}.  
In terms of the effective mass of light neutrinos, 
the above  implies \cite{epj} $|m_{ee}|< 0.35  \,\rm{eV}$. 
The experimental claim %of observation of  $0 \nu 2\beta$ 
by Klapdor and collaborators \cite{Klapdor} implies  
$|m_{ee}|=0.23 \pm 0.02 \pm 0.02 \,\rm{eV}$ \cite{mfg,Klapdor,f2010} at $68\%$ {\em{C.L}} . This experimental hint of $0 \nu 2\beta$
 challenges the result  from  cosmology 
\cite{fogli-05}, as  emphasized  in Fig.~\ref{cosmo-klap}. In the left panel, the 
effective mass $m_{ee}$ vs the lightest neutrino mass \cite{vis-mee} has been shown 
 (note the  region disfavored 
 from cosmology), while 
the figure in the right panel (note the linear scale) shows the combination of   Klapdor's claim and the 
recent cosmological \cite{sloan}  bound $\Sigma m_i <0.26 \, \rm{eV}$ at $95\%$ {\em{C.L}}. It is evident from the figure, 
 that the light neutrino contribution can not reach the Klapdor's limit \cite{Klapdor}, 
if we  consider the cosmological bound seriously. 
The above discussion suggests  us to think for an alternative possibility:  whether  any  new contribution to  $0\nu 2\beta$   
can be large enough to saturate the  experimental bound (or hint).

\section{Heavy Sterile Neutrino contribution}
\begin{figure}[htb]
\begin{center}
\begin{minipage}[h]{0.3\linewidth}
\vspace*{1cm}
\caption{Bounds on the mixing between the electron neutrino and a (single) 
sterile  neutrino as obtained from Eq.~\ref{sa}. 
For comparison, 
we also show other experimental constraints
as compiled in Atre {\it et al.}. See text for  details. \label{bound}}
\end{minipage}
\hspace*{1cm}
\begin{minipage}[t]{0.5\linewidth}
\vspace*{-2.5cm}
\includegraphics[width=0.73\textwidth, angle=270]{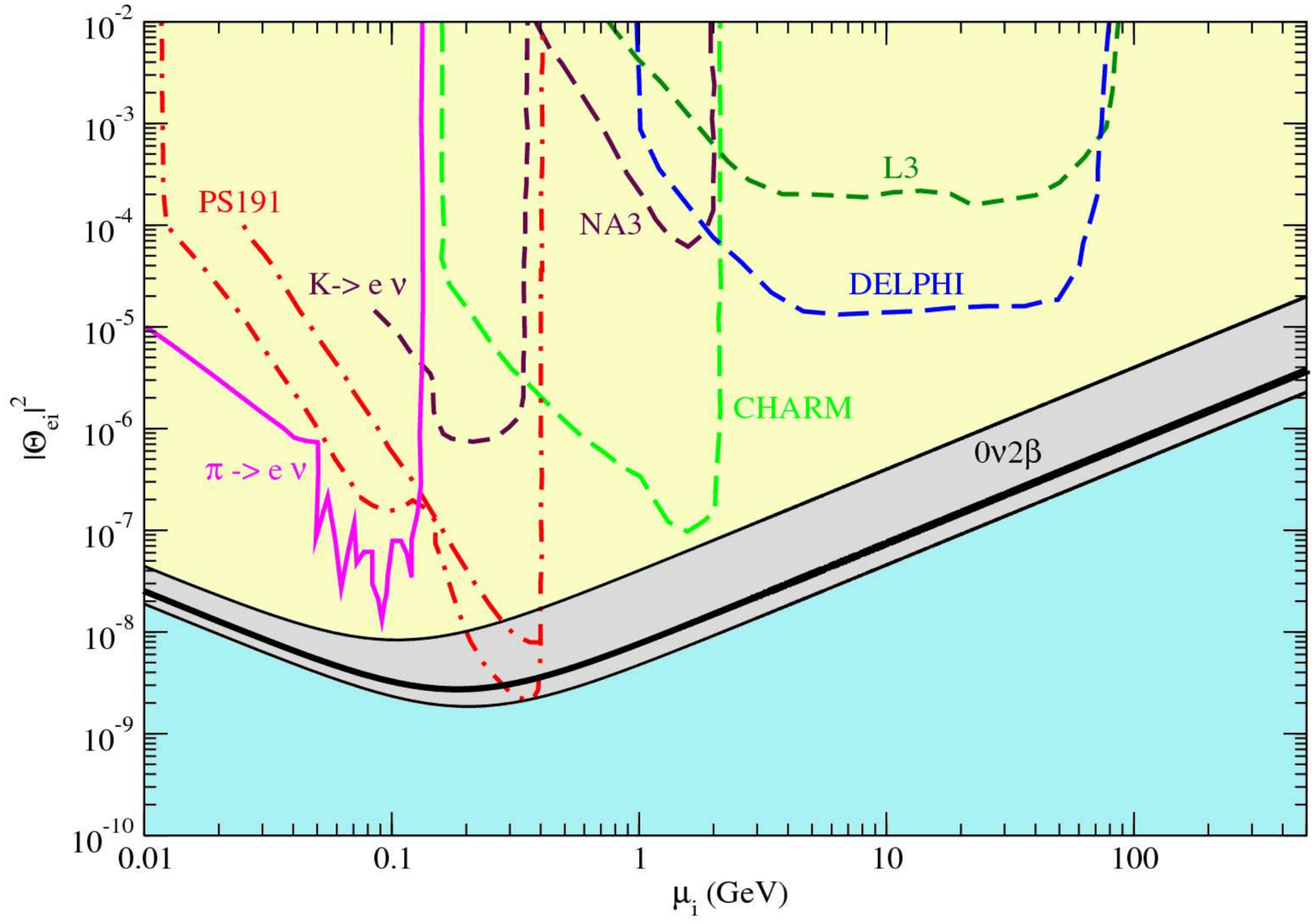}
\end{minipage}
\end{center}
\end{figure}

Consider $n_h$ generation of heavy sterile neutrinos with mass $M_i$ and 
mixing $V_{ei}$. Let us define $\eta_\nu=U_{ei}^2 m_i/m_e$ 
and $\eta_N=V_{ei}^2 m_p/M_i$. The traditional expression of the half-life is:
\bea
\frac{1}{T_{1/2}}=G_{0\nu} \left| \mathcal{M}_\nu \eta_\nu +\mathcal{M}_N \eta_N \right|^2,
\label{lif}
\eea
where
$\mathcal{M}_{\nu}$ and $\mathcal{M}_N$ are 
the nuclear matrix elements for light and heavy exchange respectively, $G_{0 \nu}$ is the phase-space factor \cite{f2010}. One 
can recast this into a useful form \cite{kovalenko}:
\begin{equation}\label{sa}
\frac{1}{T_{1/2}}=K_{0\nu} \left| \Theta_{ei}^2 \frac{ \mu_i}{\langle p^2 \rangle-\mu_i^2}  \right|^2.
\end{equation}
where
$K_{0\nu}=G_{0\nu} ({\mathcal{M}_N}m_p)^2$
and $
\langle p^2 \rangle\equiv -m_e m_p \frac{\mathcal{M}_N}{\mathcal{M}_\nu}
$. This agrees with the Eq.~\ref{lif}, if one identifies,  
$
(\mu_i,\ \Theta_{ei})=(m_i,\ U_{ei})$ for $\mu_i\to 0$ and $
(M_i,\ V_{ei}) $ for $\mu_i\to \infty$. The scale of comparison is  
$\langle p^2 \rangle \sim (200)^2$ $\rm{MeV}^2$, the typical size of Fermi momentum inside the nucleus. Using Eq.~\ref{sa}, 
we  obtain   the bound
on the mass and mixing parameters, shown in  Fig.~\ref{bound}. In addition,  we  also show  the 
bounds coming from different meson decay experiments as well as heavy neutrino decay experiments \cite{atre} for comparison
\footnote{For more detail on the  bounds from meson decay and heavy neutrino decay, see \cite{atre}.
Also, to compare the bounds coming 
from lepton number violating $B^{-}$  meson decays, see Aaij {\it et al.} \cite{LHCb}}.
The upper yellow region in Fig.~\ref{bound} is 
disfavored by neutrinoless double beta decay experiment. 
The thick black line in the middle gray band represents the present bound on the mass and mixing angle, 
where the updated nuclear matrix elements \cite{f2010} ${\mathcal{M}_{\nu}}=5.24$ 
and ${\mathcal{M}_{N}}=363$ have been used.  In terms of numerical values, the bound corresponds  to 
$\frac{\Theta^2_{ei}}{\mu_i} \leq 7.6 \times 10^{-9}\,\, \rm{GeV}^{-1}$.  The 
upper thin black line corresponds to the previous bound \cite{atre,f2005}, while
the lower line represents rather a conservative limit. Evidently,  
the most stringent bound on active-sterile neutrino mixing  comes from neutrinoless double beta decay experiment.

\vspace*{-0.1cm}
\section{ Type I seesaw and naive expectation}
\begin{figure}[b]
%\vspace*{-0.31cm}
\begin{center}
\begin{minipage}[h]{0.3\linewidth}
%\vspace*{-3cm}
\caption{
Naive expectations on Type I seesaw model are  displayed  on 
the $(m-M)$-plane. The constraints from $0\nu 2 \beta$ transition 
heavy Majorana neutrino searches in colliders 
and lepton flavor violating decays are shown. See the text for detailed 
explanation.\label{xxx}}
\end{minipage}
\hspace*{1cm}
\begin{minipage}[h]{0.6\linewidth}
\includegraphics[width=0.75\textwidth, angle=0]{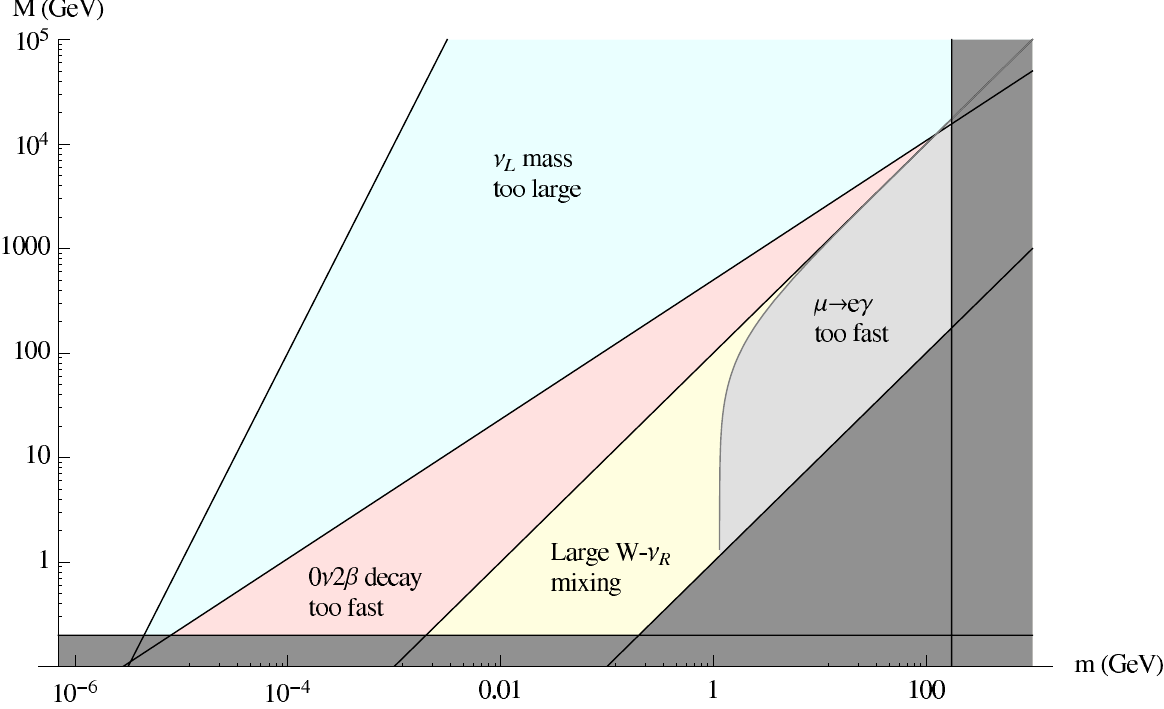}
\end{minipage}
\end{center}
\end{figure}

In this section, we consider the minimal seesaw scenario Type I seesaw \cite{seesawall}. We 
discuss what is the {\em naive expectations} from the sterile neutrino states
in a) neutrinoless double beta decay process b) heavy neutrino searches at 
colliders c)  lepton flavor violating processes. In order to discuss the above mentioned points, 
we denote the mass scale of $M_D$ and $M_R$ 
by two parameters $m$ and $M$ respectively.  The {\em naive expectations} 
from the sterile neutrino states  are  shown in Fig.~\ref{xxx}.
The details of the figure are as follows, 
\begin{itemize}

\item
 The three gray bands are excluded from the following considerations:
(a) $M>200$ MeV, i.e., 
heavy sterile  neutrinos are assumed to act as point-like 
interactions in the nucleus.
(b) $m<174$ GeV, in order to ensure perturbativity of the Yukawa couplings;
(c) $M>m$, namely,  seesaw in a conventional sense.

\item
The remaining  $(m-M)$-plane is divided in various regions by the 
three oblique lines, defined as follows:
 (1) The leftmost oblique line (between white and blue region) 
 corresponds to  neutrino mass  $M_\nu\sim m^2/M=0.1$ eV.
The region below this line is excluded from light neutrino mass constraint.
 (2) The line between pink and blue region represents the 
contribution from heavy sterile neutrinos, which saturate the Heidelberg-Moscow
bound \cite{epj}. The region below this line is excluded, since contribution 
larger than the experimental bound is achieved in this region.  
 (3) The  oblique line  that  separates the 
 pink and yellow region corresponds to  large mixing between 
active and sterile neutrinos, i.e.,  $V_{\mu i}\sim {m}/{M} \sim 10^{-2}$. 
In the region below this line,  the production of the heavy Majorana  
neutrinos in colliders is not suppressed by a small coupling \cite{aguila,smirnov}. 
(4) Finally, we also show the constraints coming from $\mu \to e\gamma$ process \cite{MEG,tommasini}. 
\end{itemize}

The above discussion clearly suggests, that the {\em naive expectations}
on Type I seesaw rules out  large contribution to $0\nu 2\beta$ 
from heavy sterile neutrinos, or the prospect of 
heavy neutrino searches at collider or even a 
 rapid $\mu\to e\gamma$ transition.

\section{ Dominant sterile contribution in multiflavor scenario}

In this section  we show how to achieve 
a dominant sterile 
neutrino contribution in neutrinoless double beta decay process. For this purpose, we go beyond naive dimensional
analysis, discussed in the previous section.
However, to achieve  dominant contribution to $ 0 \nu 2\beta$ from sterile neutrinos,  the light neutrino
masses have  to be smaller than the {\em naive seesaw expectation}.
Below, we discuss this possibility  in detail.

\subsection{Vanishing seesaw condition and the perturbation}

We are interested to the case  when the {\em naive expectation} 
for the light neutrino mass, $M_D^T M_R^{-1} M_D \sim m^2/M$, 
typical of Type I seesaw, does not hold.  For this purpose, let us start with 
the vanishing seesaw condition
$M^T_D M^{-1}_R M_D=0$. This  is compatible with an 
invertible right handed mass matrix $M_R$ and a non-trivial Dirac mass 
matrix  $M_D$, if in the Dirac-diagonal basis the two matrices have the following form \cite{mfg}: 
\bea
M_D=\pmatrix { 0 & 0 & 0 \cr 0 & 0 & 0 \cr 
0 & 0 & m }; \, \, M_R= \pmatrix{ 0 & 0 & M_1 \cr 0 & M_2   & M_3 \cr 
M_1 & M_3 & M_4} 
\eea
The light neutrino masses will be generated as a perturbation of this 
vanishing seesaw condition. For definiteness, 
we consider the following perturbation of the Dirac and Majorana 
mass matrix:
\bea
M_D =m \, \mbox{diag}(0,\epsilon,1)\ \ \
M_R^{-1}=
%\sim
M^{-1}
\left(
\begin{array}{ccc}
1 & 1 & 1 \\
1 & 1 & 1 \\
1 & 1 & \epsilon
\end{array}
\right)
\Rightarrow
M_\nu=
\frac{m^2}{M}
\left(
\begin{array}{ccc}
0 & 0 & 0 \\
0 & \epsilon^2 & \epsilon \\
0 & \epsilon & \epsilon
\end{array}
\right)
\label{casea}
\eea
where $\epsilon$ is the perturbing element and $\epsilon<1$. 
In Eq.~\ref{casea}  we have written down elements which are of  ${\mathcal{O}}(1)$. For notational 
clarity we skip writing  the 
coefficients of ${\mathcal{O}}(1)$ terms
 explicitly.
It is evident from Eq.~\ref{casea} that  the light neutrino mass matrix depends crucially on the perturbing 
element $\epsilon$,  and as  $\epsilon \to 0$, the light neutrino mass matrix becomes zero.

\subsection{ Sterile contribution in neutrinoless double beta decay}
For $M^2_i \gg |p^2| \sim (200)^2\mbox{ MeV}^2$ and for real $M_R$, the amplitude  for 
the light and heavy neutrinos are represented by $ \frac{m_{ee}}{p^2} $ and 
$(M_D^T M_R^{-3} M_D)_{ee}$ respectively. Going from Dirac-diagonal to 
the flavor basis, the sterile contribution will be $ \kappa \frac{m^2}{M^3}$
where the factor $\kappa$ contains the information of the change of basis. For the particular example
which we have discussed in the previous section,  this turns out to be: 
\be
(M_D^T M_R^{-3} M_D)_{ee}^{\mathrm{(Fl.)}}= 
\xi \frac{m^2}{M^3}\times 
\left\{
\begin{array}{cl}
\frac{ (U^*_{e2} \sqrt{m_2} +U^*_{e3} \sqrt{m_3})^2}{m_2+m_3}
& \mbox{ with normal hierarchy}\\[2ex]
\frac{(U^*_{e2} \sqrt{m_2}+ U^*_{e1} \sqrt{m_1})^2}{m_1+m_2}
& \mbox{ with inverted hierarchy}
\end{array}
\right.
\label{conthree}
\ee
where $\xi$ is an ${\mathcal{O}}(1)$ factor that depends on the elements of $M_R$.
The light neutrino contributions are
$|m_{ee}|=| m_3 U_{e3}^2-m_2 U_{e2}^2|$ 
(resp., $|m_{ee}|=| m_2 U_{e2}^2-m_1 U_{e1}^2|$) for normal (resp., inverted) mass hierarchy. Note that, 
both the numerator and denominator 
in Eq.~\ref{conthree} depends  on the light neutrino masses $m_{1,2}$ or 
$m_{1,3}$ in the same way. Hence, the sterile neutrino contributions are not suppressed from the 
smallness of light neutrino mass. Depending on the factor $\frac{m^2}{M^3}$, the sterile neutrinos
can give dominant contribution in neutrinoless double beta decay 
\footnote{
See \cite{mfg} for  discussion on another
 interesting seesaw scenario  that leads to  similar conclusion.}. 
To give an estimate, for $\frac{m^2}{M^3} \sim 
7.6 \times 10^{-9}\, \rm{GeV}^{-1}$, the sterile neutrinos can saturate the present bound \cite{epj} on $0 \nu 2\beta$ half-life.
See the texts in \cite{mfg} for the other  cases leading to similar conclusions.

\section{ Upper bound on the sterile neutrino mass scale}

In this section, we discuss what could be the upper bound on the sterile neutrinos mass scale $M$, that is consistent 
with  radiative stability. Note that, the dominant contribution in $0 \nu 2\beta$ will imply that the 
Dirac mass scale $m$ and Majorana mass scale of sterile neutrinos $M$ should be related as follows: 
\be\label{bomba}
M=16\mbox{ TeV}\times \left( \frac{T_{1/2}}{1.9\times 10^{25}\mbox{yrs}}\right)^{1/6} 
\left( \frac{\mathcal{M}_N\times \kappa}{363\times 1}\right)^{1/3} \left( \frac{m}{174\mbox{ GeV} }\right)^{2/3}
\label{16tev}
\ee
Hence, if $m$ reaches to its upper bound $174 \, \rm{GeV}$,  $T_{1/2}=1.9 \times 10^{25} \, \rm{yrs}$, and the 
nuclear matrix element is
${\mathcal{M}_N}=363$, sterile neutrinos of mass up to  $M \sim 16 \, \rm{TeV}$ can saturate the experimental bound. However, 
the question is whether the loop correction of the light neutrino masses can put further constraints on this mass scale.
Note that in this case,  when  $M \sim 16 \, \rm{TeV}$ and $m \sim 174 \, \rm{GeV}$, one will need excessive 
fine-tuning $\epsilon \sim 10^{-9}$ to satisfy the neutrino mass 
constraint $\epsilon \frac{m^2}{M} <0.1 \, \rm{eV}$. Decreasing $M$ as well as $m$ from their maximum values will however 
reduce the fine-tuning \cite{mfg}. 

The one loop correction to the light neutrino mass is \cite{smirnov} $
\delta M_\nu\sim \frac{g^2}{(4\pi)^2} \frac{m^2}{M} \log(M_1/M_2)$, if $M$ is larger than electroweak scale. On the 
other hand, for $M$ smaller than electroweak scale, the loop correction has a polynomial nature, i.e., 
$
\delta M_\nu\sim \frac{g^2}{(4\pi)^2} \frac{m^2}{M}  \frac{M^2}{M_{ew}^2}
$. If combined with the following two considerations: a) smallness of light neutrino mass $\epsilon \frac{m^2}{M}< 0.1 \, \rm{eV}$
and b) sterile neutrinos saturating  the present bound on $0\nu 2\beta$ half-life, the sterile neutrino mass scale 
turns out to be smaller than 10 GeV. See \cite{mfg} for  a  detailed discussion. 

\section{Conclusion}

Neutrinoless double beta decay is a major experiment to probe lepton number
violation. On the experimental side, there is scope of order 
of magnitude improvement  of the half-life of this process. 
We have considered the most basic Type I seesaw scenario and studied the 
sterile neutrino contribution in detail. We find that due to improvement of
nuclear matrix elements, the bound on active-sterile mixing angle is now
improved by one order of magnitude. Despite of the naive expectations that
the light neutrinos give dominant contribution, heavy sterile neutrinos can saturate the present experimental bound. 

\section*{Acknowledgments}
M.M would like to thank the organisers of  EW Interaction  and Unified Theories, Moriond 2012 for their invitation. Special thanks to Pilar Hernandez and 
Jean-Marie Fr\`ere.


\footnotesize

\section*{References}
\begin{thebibliography}{99}

\bibitem{epj}
H.~V.~Klapdor-Kleingrothaus {\it et al.},
  Eur.\ Phys.\ J.\  {\bf A12 } (2001)  147-154. 

\bibitem{expp}
 I.~Abt {\it et al.},
  [hep-ex/0404039]; S.~Schonert {\it et al.} [GERDA Collaboration],
   Nucl.\ Phys.\ Proc.\ Suppl.\  {\bf 145}, 242-245 (2005);
C.~Arnaboldi {\it et al.} [CUORE Collaboration],
 Nucl.\ Instrum.\ Meth.\  {\bf A518}, 775-798 (2004);
%
%\bibitem{expf}
  R.~Arnold {\it et al.} [SuperNEMO Collaboration],
    Eur.\ Phys.\ J.\  {\bf C70}, 927-943 (2010);
  V.~E.~Guiseppe {\it et al.} [Majorana Collaboration]
    [arXiv:0811.2446 [nucl-ex]].
%
%
\bibitem{mfg}
  M.~Mitra, G.~Senjanovic and F.~Vissani,
  Nucl.\ Phys.\ B {\bf 856} (2012) 26.

%
\bibitem{werner-rev}
  W.~Rodejohann,
  [arXiv:1106.1334 [hep-ph]].
%

\bibitem{bookk}
{\em Seventy 
years of double beta decay: From nuclear physics to beyond-standard-model particle physics,}
H.~V.~Klapdor-Kleingrothaus, Hackensack, USA. World Scientific (2010).
  %
\bibitem{zuber}
 K.~Zuber,
  %``Neutrinoless double beta decay experiments,''
  Acta Phys.\ Polon.\  {\bf B37 } (2006)  1905.
%
\bibitem{visreview}
  A.~Strumia and F.~Vissani,
  %``Neutrino masses and mixings and..,''
  arXiv:hep-ph/0606054.
%
\bibitem{rev-ponte}
B.~M.~Pontecorvo,
  %``Pages In The Development Of Neutrino Physics,''
  Sov.\ Phys.\ Usp.\  {\bf 26 } (1983)  1087-1108.
%
\bibitem{review-osc-petcov}
 S.~M.~Bilenky, S.~T.~Petcov,
  %``Massive Neutrinos and Neutrino Oscillations,''
  Rev.\ Mod.\ Phys.\  {\bf 59}, 671 (1987).
%
\bibitem{mik-shap}
S.~P.~Mikheyev, A.~Y.~Smirnov,
  %``Resonant neutrino oscillations in matter,''
  Prog.\ Part.\ Nucl.\ Phys.\  {\bf 23 } (1989)  41-136.
%
\bibitem{gelmini-rev}
 G.~Gelmini, E.~Roulet,
  %``Neutrino masses,''
  Rept.\ Prog.\ Phys.\  {\bf 58}, 1207-1266 (1995).
%
\bibitem{kail}
 K.~ Zuber, 
  %``On the physics of massive neutrinos,''
  Phys.\ Rept.\  {\bf 305 } (1998)  295.
  %
\bibitem{nir-garcia-rev}
M.~C.~Gonzalez-Garcia, Y.~Nir,
  %``Neutrino masses and mixing: Evidence and implications,''
  Rev.\ Mod.\ Phys.\  {\bf 75}, 345-402 (2003).
%
\bibitem{rev-moha-smir}
R.~N.~Mohapatra, A.~Y.~Smirnov,
  %``Neutrino Mass and New Physics,''
  Ann.\ Rev.\ Nucl.\ Part.\ Sci.\  {\bf 56 } (2006)  569-628.
 %%
%  
\bibitem{Senjanovic:2011zz}
  G.~Senjanovi\'c,
  %``Neutrino mass: From LHC to grand unification,''
  Riv.\ Nuovo Cim.\  {\bf 034}, 1-68 (2011).

\bibitem{Klapdor}
 H.~V.~Klapdor-Kleingrothaus {\it et al.},
    Phys.\ Lett.\  {\bf B586}, 198-212 (2004); H.~V.~Klapdor-Kleingrothaus {\it et al.},  Mod.\ Phys.\ Lett.\  {\bf A21}, 1547-1566 (2006).
% 
\bibitem{0nu2beta-old}
 G.~Racah,
    Nuovo Cim.\  {\bf 14}, 322-328 (1937);
  W.~H.~Furry,
    Phys.\ Rev.\  {\bf 56}, 1184-1193 (1939).
  %%
\bibitem{Majorana}
 E.~Majorana,
  Nuovo Cim.\  {\bf 14}, 171-184 (1937).
  %% 
\bibitem{feinberg}
G.~Feinberg, M.~Goldhaber, Proc.\ Nat.\ Ac.\ Sci.\ USA\ {\bf 45}, 
1301  (1959); B.~Pontecorvo,  Phys.\ Lett.\  {\bf B26}, 630-632 (1968).
  %
\bibitem{ms0nu2beta}
  R.~N.~Mohapatra, G.~Senjanovi\'c,
    Phys.\ Rev.\  {\bf D23 } (1981)  165;
 R.~N.~Mohapatra,
 Phys.\ Rev.\  {\bf D34}, 3457-3461 (1986);
 K.~S.~Babu, R.~N.~Mohapatra,
  Phys.\ Rev.\ Lett.\  {\bf 75}, 2276-2279 (1995).
%
\bibitem{pion-ex}
J.~D.~Vergados, Phys. Rev. D 25, 914 917 (1982);
  S.~Bergmann {\it et al.},
  Phys.\ Rev.\  {\bf D62}, 113002 (2000); A.~Faessler {\it et al.},
   Phys.\ Rev.\  {\bf D77}, 113012 (2008);   M.~Hirsch {\it et al.},
  Phys.\ Lett.\  {\bf B352}, 1-7 (1995). 
%%
\bibitem{tello}
V.~Tello {\it et al.},
  Phys.\ Rev.\ Lett.\  {\bf 106}, 151801 (2011);  
  M.~Nemevsek {\it et al.},
  arXiv:1112.3061 [hep-ph].
%
\bibitem{ibarra-blenow}
 A.~Ibarra {\it et al.},
  JHEP {\bf 1009}, 108 (2010); 
  M.~Blennow {\it et al.},
  JHEP {\bf 1007 } (2010)  096.
%
 \bibitem{choubey-sruba}
 S.~Choubey {\it et al.}, JHEP 05(2012)017
  %[arXiv:1201.3031 [hep-ph]]; 
%; 
J.~Chakrabortty {\it et al.},
arXiv:1204.2527 [hep-ph].
%
\bibitem{seesawall}  
 P.~Minkowski,
   Phys.\ Lett.\  {\bf B67}, 421 (1977); R.~N.~Mohapatra, G.~Senjanovi\'c,
   Phys.\ Rev.\ Lett.\  {\bf 44}, 912 (1980);
 T.~T.~Yanagida, in {\it Proceedings of the Workshop on the Unified Theory and the Baryon Number in the Universe} (O.Sawada and A.Sugamoto, eds.), KEK, Tsukuba, Japan, 1979, p.95; 
 M.~Gell-Mann, P.~Ramond, R.~Slansky, Supergravity (P. van Nieuwenhuizen et al.eds), North Holland, Amsterdam, 1980.
%
\bibitem{f2010}
  F.~\v{S}imkovic, J.~Vergados, A.~Faessler,
  Phys.\ Rev.\  {\bf D82}, 113015 (2010).
%%
\bibitem{fogli-05}
 G.~L.~Fogli {\it et al.},
   Phys.\ Rev.\  {\bf D78}, 033010 (2008).
%    

\bibitem{vis-mee}
 F.~Vissani,
   JHEP {\bf 9906}, 022 (1999).
%

\bibitem{sloan}
 R.~de Putter {\it et al.},
   arXiv:1201.1909 [astro-ph.CO].
 %
\bibitem{kovalenko}
  S.~Kovalenko, Z.~Lu, I.~Schmidt,
  Phys.\ Rev.\  {\bf D80 } (2009)  073014.
%
\bibitem{atre}
   A.~Atre, T.~Han, S.~Pascoli, B.~Zhang,
  JHEP {\bf 0905 } (2009)  030.
   %
%%
\bibitem{f2005}
P.~Benes, A.~Faessler, F.~\v{S}imkovic, S.~Kovalenko,
  Phys.\ Rev.\  {\bf D71 } (2005)  077901.
%
\bibitem{LHCb}
R.~Aaij {\it et al.}  [LHCb Collaboration],
  %``Searches for Majorana neutrinos in B- decays,''
  arXiv:1201.5600 [hep-ex].
  %%CITATION = ARXIV:1201.5600;%%

%
\bibitem{aguila}
F.~del Aguila, J.~A.~Aguilar-Saavedra,
  Phys.\ Lett.\  {\bf B672}, 158-165 (2009);
F.~del Aguila, J.~A.~Aguilar-Saavedra,
  Nucl.\ Phys.\  {\bf B813}, 22-90 (2009).
%
\bibitem{smirnov}
  J.~Kersten, A.~Y.~Smirnov,
  Phys.\ Rev.\  {\bf D76}, 073005 (2007).
%
\bibitem{MEG}
 J.~Adam {\it et al.}  [MEG Collaboration],
  Nucl.\ Phys.\ B {\bf 834} (2010) 1-12 (2010).
%
\bibitem{tommasini}
 D.~Tommasini {\it et al.},
  Nucl.\ Phys.\  {\bf B444}, 451-467 (1995).
 %

\end{thebibliography}
\end{document}